\documentclass[aps, prl,  twocolumn,preprintnumbers,amsmath,amssymb,floatfix]{revtex4}

\usepackage{graphicx}
\usepackage{epstopdf}
\usepackage{dcolumn}
\usepackage{bm}

\begin{document}

\title{Confined space and effective interactions of multiple self-avoiding chains}

\author{Suckjoon Jun}
 \altaffiliation{Present address: Facult\'{e} de M\'{e}decine, INSERM Site Necker, U571, 156 rue de Vaugirard, 75015 Paris, France (email: suckjoon.jun@necker.fr)}
  \author{Axel Arnold}
   \affiliation{FOM-Institute AMOLF, Kruislaan 407, 1098 SJ Amsterdam, the Netherlands}
\author{Bae-Yeun Ha}
 \affiliation{Department of Physics and Astronomy, University of Waterloo, Waterloo, Ontario N2L 3G1, Canada}

\date{\today}

\begin{abstract}

We study the link between three seeming-disparate cases of self-avoiding polymers: strongly overlapping multiple chains in dilute solution, chains under spherical confinement, and the onset of semi-dilute solutions.  Our main result is that the free energy for overlapping $n$ chains is \emph{independent} of chain length and scales as $n^{9/4}$, slowly crossing over to $n^3$, as $n$ increases. For strongly confined polymers inside a spherical cavity, we show that rearranging the chains does not cost an additional free energy. Our results imply that, during cell cycle, \emph{global} reorganization of eukaryotic chromosomes in a large cell nucleus could be readily achieved.

\end{abstract}


\maketitle

How are polymers organized inside a confined space?  What is the free energy barrier to the overlapping of two or more chains in the absence or presence of confinement?  Is the barrier higher for longer chains? Recently, there has been renewed interest in the problem of confined polymers~\cite{Cremer04, Morrison, Sakaue, Cacciuto}, because of its relevance to such biological processes as DNA packaging in a virus~\cite{Gelbart} and organization and segregation of chromosomes in bacteria~\cite{Jun}. In eukaryotes, multiple chromosomes are encapsulated in a cell nucleus, \i.e., in a dimension many times smaller than their natural sizes. Although a single cell can contain as many as billions of basepairs of DNA (e.g., human), the dimension $\agt$ 10 $\mu$m of cell nuclei implies that the volume fraction of total amount of DNA is typically much less than that of virus or bacteria, and the eukaryotic chromosomes in a spherical volume may be considered as a semi-dilute polymer solution~\cite{Cremer04}. Here, one of the key issues is spatial organization of chromosomes. The emerging view is that they are compartmentalized and occupy discrete ``territories'' inside a nucleus~\cite{Cremer}. However, chromosomes should also be able to mix, when necessary (e.g., recombination), and, indeed, they do~\cite{Pombo}. Unfortunately, despite its importance, little is known about how self-avoiding polymers interact and are organized in a confined space, whereas the effective interaction of polymers in dilute (bulk) solution has been well studied~\cite{Likos}.  Although these two subjects, namely, the effective interaction in \emph{dilute} solution and polymers under \emph{confinement}, have been regarded a disparate grouping so far, as we shall show below, there is a close connection between the two.

The first attempt to characterize the effective interaction between two polymeric coils in dilute solution, each carrying $N$ monomers, was made as early as in 1950 by Flory and Krigbaum~\cite{Flory}.
The main conclusion of their ``mean-field" approach is that the overlapping free energy, {\i.e.}, the free energy cost for bringing two chains in a volume explored by each chain, scales as $\beta \mathcal{F}_\mathrm{FK} \sim N^{\frac{1}{5}} \gg 1$ (Throughout this paper, $\beta =1/k_BT$, where  $k_B$ is the Boltzmann constant and $T$ the absolute temperature.) 
The repulsion between long chains is so strong that they should behave as mutually impenetrable hard spheres. 
 Partly due to its simplicity, this picture satisfied scientists for about three decades. Then, the more careful scaling analysis of Grosberg \emph{et al.}~\cite{Grosberg}, which takes into account monomer-density correlations~\cite{Cloizeaux,Daoud}, showed that the overlapping free energy is of the order of $k_BT$ and is asymptotically {\it independent} of $N$. This counter-intuitive result has since been confirmed by renormalization group calculations~\cite{Kruger} and numerical simulations~\cite{Hall, Louis}.   

\begin{figure}[t]
\centering
\includegraphics[width=3.4in]{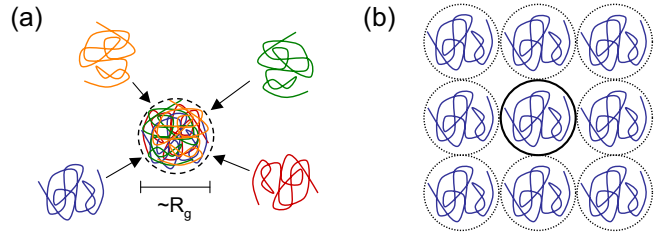}
\caption{(Color online) (a) Strongly overlapping multiple chains with excluded volume in dilute solution. (b) Onset of semi-dilute regime as ``stacking spheres.''  Above the onset concentration, each sphere is independent of one another.   This allows one to map the overlapping chains in (a) onto those confined in a single sphere in (b) under equivalent conditions --  the same chain (sphere) size and the same monomer concentration -- and thus to estimate the overlapping free energy in (a).}
\label{fig:onset}
\end{figure}

\begin{figure}[t]
\centering
\includegraphics[width=3.4in]{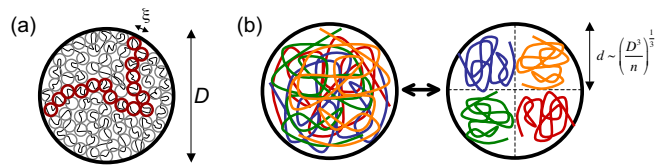}
\caption{(Color online) A self-avoiding chain under spherical confinement.
(a) Stacking blobs in a moderately-strongly confined self-avoiding chain, {\i.e.}, $D \gg \xi \gg a$.
(b) Mixing and de-mixing of chains within a confined sphere.}
\label{fig:stacking_blobs}
\end{figure}

The purpose of this Letter is to unravel the link between a single chain under spherical confinement, a system of multiple, strongly overlapping chains in dilute solution, and the onset of semidilute solution [Figs.~\ref{fig:onset} and~\ref{fig:stacking_blobs}], using scaling arguments and molecular dynamics simulations.   The polymers we consider here are flexible chains with excluded volume, unless otherwise stated. An important corollary of our analysis is a non-trivial generalization of the aforementioned two-chain result by Grosberg \emph{et al.} to the case of an arbitrary number of chains [see Fig.~\ref{fig:onset}(a)] in a wide range of monomer concentrations.  It also illuminates the weak dependence of the effective interaction on polymer concentration studied recently by Louis \emph{et al.}~\cite{Louis} using numerical simulations.   Biological implications of our results will also be briefly discussed.

Consider a polymer solution, where  its constituent chains start to touch one another [Fig.~\ref{fig:onset}(b)].  At this \emph{onset} of the so-called semi-dilute regime, the solution can be viewed as stacking of imaginary spheres, where the size of each sphere, containing a single chain, is $\sim$$R_g$, {\i.e.}, the radius of gyration of each chain in the dilute regime.  Using this ``stacking sphere'' picture, we can, in fact, apply the physics of {\em semidilute} solutions to estimate the interaction free energy of multiple chains in {\em dilute} solution [Fig.~\ref{fig:onset}(a)].  The essence of our analysis is that, as the monomer concentration increases above the onset concentration, each sphere shown in Fig.~\ref{fig:onset}(b) behaves as an independent replica of the neighboring ones (see below).   

To this end, we first consider a long chain formed by $N$ monomers, compressed inside a spherical cavity of diameter $D < R_g$.  In the case of an ideal chain, it is well known that the confinement free energy can be obtained by a random-walk (RW) analysis, where the chain can be divided into ``independent" subchains that start from one point on the wall and reach another via a random walk~\cite{GrosbergBook}.  This is reasonable since, at each ``collision" on the wall, the chain loses its memory of chain connectivity~\cite{Cassasa}.  Since the length of a subchain scales as $N_\mathrm{sub} \sim (D/a)^2$, where $a$ is the monomer size, the confinement  free energy 
({\i.e.}, entropy loss of order $1$ times $-k_BT$) can be straightforwardly obtained by counting the total number of independent subchains as follows~\cite{GrosbergBook}
\begin{equation}
\label{eq:ideal}
\beta \mathcal{F}_\mathrm{RW} \sim \frac{N}{N_\mathrm{sub}} \sim \bigg(\frac{R_g}{D}\bigg)^2.
\end{equation}

In the case of a strongly confined chain with excluded volume [or a confined ``self-avoiding walk (SAW)''] [Fig.~\ref{fig:stacking_blobs}(a)], the major source of increase of free energy is the collisions between monomers along the chain.  (This allows us to approximate the spheres in Fig.~\ref{fig:stacking_blobs}(b) as independent subsystems.)  The confinement free energy can be obtained by mapping the chain onto an equivalent semidilute solution:  In the semidilute regime, this monomer-monomer contact probability within the sphere of volume $V \simeq D^3$ can be estimated using the des Cloizeaux exponent $1/(3\nu-1)$~\cite{Cloizeaux}. The resulting free energy is 
\begin{equation}
\label{eq:SAW}
\beta \mathcal{F}_\mathrm{SAW} \simeq N  \bigg(\frac{Na^3}{V}\bigg)^\frac{1}{3\nu-1}
\simeq \bigg(\frac{R_g}{D}\bigg)^\frac{3}{3\nu-1},
\end{equation}
where $\nu \simeq 3/5$ is the Flory exponent~\cite{Grosberg, GrosbergBook}. Note that, as expected, $\mathcal{F}_\mathrm{SAW}$ grows with $N$ faster than linearly, namely, faster than that of the corresponding ideal chain (Eq.~\ref{eq:ideal}). Although useful to estimate the free energy, this particle picture provides little insight into the spatial organization of the confined chain. A more intuitive approach is that of stacking ``blobs'' [Fig.~\ref{fig:stacking_blobs}(a)]~\cite{Abadie, deGennes, Sakaue}, where the size of each blob is the correlation length $\xi \sim a(D^3/a^3N)^{\nu/(3\nu-1)}$ in an equivalent semidilute solution. We note that Eq.~\ref{eq:SAW} is then self-consistently restored by the ``$k_BT$ per blob" ansatz, {\i.e.}, $\beta \mathcal{F}_\mathrm{SAW} \sim D^3/\xi^3$~\cite{Sakaue}. 

Within the stacking-blob picture (not to be confused with the \emph{stacking-sphere} picture), we can view a long chain with excluded volume confined in a spherical cavity as a system of overlapping multiple chains. The basic idea is similar to that of the RW analysis described above: Each ``independent'' subchain is a series of connected blobs~\cite{Abadie}, which percolates and connects two points on the confining walls as illustrated by the red blobs in Fig.~\ref{fig:stacking_blobs}(a). The number of monomers per independent subchain, $N_\mathrm{sub}$, can be obtained as follows.\\

(i) Weakly confined regime ($D \agt \xi \gg a$): In polymer solution, this is analogous to the onset of the semidilute regime, where the chains start to contact one another:
$\xi \simeq R_g \simeq D$, and, thus, $N_\mathrm{sub} \simeq (D/a)^{1/\nu}$ [Fig.~\ref{fig:onset}(b)].  

(ii) Moderately-strongly confined regime ($D \gg \xi \gg a$): This is where both the semidilute regime and the stacking-blob picture in Fig.~\ref{fig:stacking_blobs}(a) apply.   In other words, within the correlation length $\xi$, the chain conformation is that of SAW, but the global conformation of each independent chain is described by the RW of blobs. Thus, $N_\mathrm{sub} \simeq (\xi/a)^{1/\nu} (D/a)^2 \simeq (D/a)^{1 / (3 \nu-1)} N^{(2 \nu-1)/(3 \nu -1)}$.

(iii) Concentrated regime ($D \gg \xi \simeq a$):
In this regime, excluded volume is screened at all length scales beyond $a$~\cite{deGennes}. The RW analysis then leads to $N_\mathrm{sub} \simeq (D/a)^2$.\\

Using the expressions of $N_\mathrm{sub}$ in (i) and (ii) (the two regimes of our main interest), we can rewrite Eq.~\ref{eq:SAW} in terms of the number of independent subchains, $n = N/N_\mathrm{sub}$, to estimate the free energy cost $\mathcal{F}_\mathrm{n}$ for overlapping an arbitrary number ($n$) of chains. We find
\begin{eqnarray}
\label{eq:overlap1}
\beta \mathcal{F}_\mathrm{n} &\simeq& n^\frac{3\nu}{3\nu-1} \simeq n^\frac{9}{4} ~~~(D \agt \xi \gg a)\\
\label{eq:overlap2}
&\simeq& n^3 ~~~~~~~~~~~~~~~(D \gg \xi \gg a).
\end{eqnarray}
In fact, this is also the free energy cost for bringing $n$ self-avoiding chains (each carrying $N$ monomers) in dilute solution to an imaginary sphere of volume $v_g$ each chain would explore otherwise [Fig.~\ref{fig:onset}(a)]. Here, $v_g \sim R_g^3 \sim (a N^\nu )^3$, a condition compatible with (i). In higher monomer concentrations compatible with (ii), the chain size is reduced by ``screening" effects~\cite{deGennes}, and, thus,
$v_g \sim (a~n^{1-3\nu} N^\nu)^3$, which explains the larger exponent in Eq.~\ref{eq:overlap2}. This rapid increase of $\mathcal{F}_\mathrm{n}$ with $n$, as evidenced in our blob picture, implies a strong repulsion between two spheres of dense chains, reminiscent of that between two star polymers (see Discussion for interesting consequences on chromosome organization).

At first glance, this $N_\mathrm{sub}$-independence is surprising -- the increase of free energy due to chain overlapping is \emph{independent} of the chain length, where its special case for $n=2$ explains the results by Grosberg \emph{et al.}~\cite{Grosberg}. This is a natural consequence of the functional form of Eq.~\ref{eq:SAW}. Furthermore, the exponent $9/4$ is identical to the osmotic pressure exponent in the semidilute regime.  This is not accidental: The monomer density (times $a^3$) is now translated into $n$, which should correctly reflect the stronger monomer-density correlation in the semidilute regime~\cite{Daoud,deGennes}. 

Importantly, the scaling form in Eq.~\ref{eq:SAW} also implies that the overlapping free energy is invariant under rearrangement of the chains as illustrated in Fig.~\ref{fig:stacking_blobs}(b), where  each segregated chain occupies a smaller volume of linear dimension $d \sim (D^3/n)^{1/3}$.  The total free energy after segregation is the same as before:
\begin{equation}
\label{eq:segregation}
\beta \mathcal{F}_\mathrm{seg} \sim n \bigg(\frac{r_g}{d}\bigg)^\frac{3}{3\nu-1} \sim \bigg(\frac{R_g}{D}\bigg)^\frac{3}{3\nu-1} \sim \beta \mathcal{F}_\mathrm{SAW},
\end{equation}
where $r_g \sim aN_\mathrm{sub}^\nu$.

Also, note that Eqs.~\ref{eq:overlap1} and~\ref{eq:overlap2} can explain the recent simulation results by Louis \emph{et al.}~\cite{Louis}. Their main conclusion is that the two-chain interaction in a polymer solution is $\approx 2 k_BT$ and independent of the monomer concentration $c$ for c up to the overlap concentration $c^*$, increasing only slightly with $c$ above $c^*$. This weak $c$ dependence is already apparent from Eq.~\ref{eq:overlap1} and~\ref{eq:overlap2}.  In the scaling regime, the ratio of entropy loss in the dilute regime and the semidilute regime is only $2^3/2^{9/4} \approx 1.7$, and, thus, we predict the two-chain interaction increases only up to 3-4 $k_BT$ from $\approx 2k_BT$.  

To augment our scaling analysis, we also performed Molecular Dynamics simulations using ESPResSo~\cite{espresso04a}. In our simulations, the polymer chains are represented as a bead-spring model. We chose a chain consisting of $N= 1000$ beads connected by FENE (spring) bonds with a purely repulsive Lennard-Jones (LJ) potential for excluded-volume interactions (monomer-monomer as well as monomer-wall). The basic length scale in the simulation is the bead diameter $a$, energies are measured in units of the LJ interaction energy $\epsilon$ at distance $a$, and the mass of a bead $m$ is the mass unit. Consequently, time is measured in units of the LJ time $\tau_{LJ}=a\sqrt{m/\epsilon}$. The FENE bond constant was chosen to
be $10\epsilon$ and the maximal elongation of a bond as $2a$. We used a standard velocity Verlet algorithm with a timestep of $0.01\tau_{LJ}$ to propagate the system, and a Langevin thermostat with friction constant $\gamma=\tau_{LJ}^{-1}$ to keep the system at constant temperature $T=\epsilon/k_B$.

We ran 40 sets of simulations to test a wide degree of confinement, namely, the radius of confining sphere $R = D/2$ lies in the range $R_g/4 \leq R \leq R_g$ (where $R_g \approx 28.3$ for $N=1000$). We started each set of simulations by generating a SAW in a large spherical cavity whose size was several times the $R_g$ of the chain. Then, the size of the confining sphere was gradually reduced, forcing the chain radially inward, until it reached the target value of $D$ ($< R_g$). Thereafter, we recorded $2 \times 10^4$ conformations and associated parameters at every $10 \tau_\mathrm{LJ}$.  
To check the consistency of our simulations, we first computed the pressure, $p$, of the confined chain, which is related to the free energy of confinement by $p=-\frac{\partial{\mathcal{F}_\mathrm{SAW}}}{\partial{V}}$ or $p \sim \phi \mathcal{F}_\mathrm{SAW} / N$, where $V \simeq D^3$ is the volume of sphere and $\phi \simeq Na^3/D^3$ the volume fraction of the chain. Since Eq.~\ref{eq:SAW} can also be expressed in terms of $\phi$ as $\mathcal{F}_\mathrm{SAW} \sim N \phi^{1/(3\nu-1)}$, we obtain an equivalent scaling for the pressure $p \sim \phi^{3\nu/(3\nu-1)} \simeq \phi^{9/4}$. This relation is indeed confirmed by our simulations (data not shown),
which lead to $p \sim \phi^{2.27 \pm 0.02}$ for $\phi < 0.75$ and are in excellent agreement with the recent Monte Carlo simulations by Cacciuto and Luijten~\cite{Cacciuto}. 
\begin{figure}[t]
\centering
\includegraphics[width=3.0in]{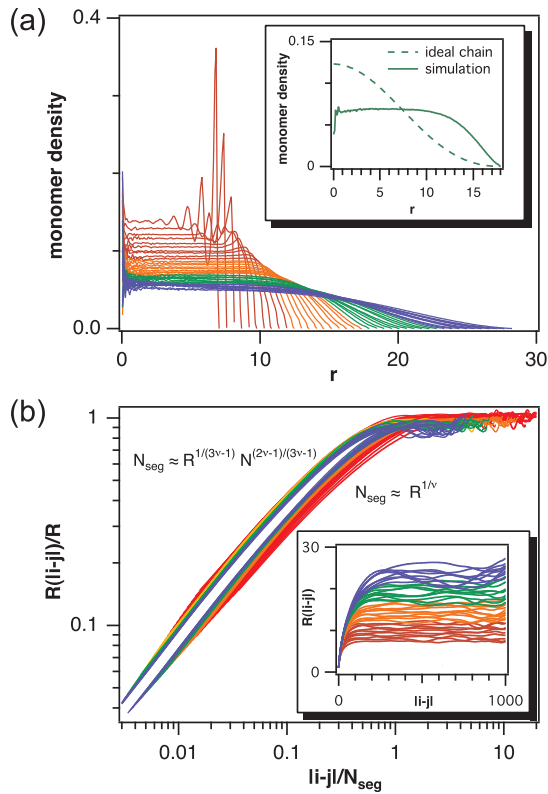}
\caption{(Color online)  Independent-subchain analysis.  (a) Monomer densities tend to be uniform for a sizable range of $|i-j|$. The liquid-like oscillatory behavior near the wall at high monomer concentrations ($\phi \agt 0.4$) illustrates how the wall cooperates in enhancing the ordering.  (b) $R(|i-j|)$ vs. $|i-j|$. The inset shows the internal distance $R(|i-j|)$ becomes saturated beyond $|i-j| \approx N_\mathrm{sub}$, due to the ``reflecting" wall.   At low concentrations (weakly confined chains, {\i.e.}, $\xi \sim D$, represented by blue and green curves), the data tend to collapse when $|i-j|$ is rescaled by $N_\mathrm{sub}=(R/a)^{1/\nu}$. For the strongly confined case $D \gg \xi \gg a$, however, the correct rescaling factor is (ii) $N_\mathrm{sub}=(R/a)^{1 / (3 \nu-1)} N^{(2 \nu-1)/(3 \nu -1)}$. These results support the independent-chain assumption.}
\label{fig:monomer_density}
\end{figure}

Next, we tested the validity of the stacking blob hypothesis illustrated in Fig.~\ref{fig:stacking_blobs}(a). If the blobs can be considered as impenetrable hard spheres of diameter $\xi$ stacked together, their radial monomer density profile in the spherical cavity tends to be uniform, except within a length of order $\xi$ from the wall. Our simulation confirms this: In Fig.~\ref{fig:monomer_density}(a), we show 40 radial density profiles, $\rho(r)$, from our simulations, which have been normalized such that $\int_0^R \rho(r) dr = 1$. Note that $\rho(r)$ is indeed constant for a significant range of $r$. This feature is particularly pronounced when we compare it with that of an ideal chain $\rho_{{}_\mathrm{id}}(r)=(1/2 \pi R) \sin^2(\pi r/R)/r^2$ [see Fig.~\ref{fig:monomer_density}(a) inset for the case of $R = 18.0 = 0.64R_g$].  The decay near $r=R$ manifests the existence of a depletion layer of length scale $\xi$ from the confining wall. One interesting observation is that, at high volume fraction ($\phi \agt 0.15$), the density becomes oscillatory near the wall. This means that the wall collaborates in enhancing the ordering --
it is the signature of the crossover from the semidilute regime ($\xi \gg a$) to the concentrated regime ($\xi \simeq a$) mentioned above. Indeed, we simulated a system of $N=1000$ hardspheres by removing the bonds between monomers in our simulations, and observed the same oscillations in $\rho(r)$ (data not shown in Fig.~\ref{fig:monomer_density}).

Our final and the most important test concerns the view of a single confined chain with excluded volume as significantly overlapping multiple, independent subchains.  If this view is correct, the average internal distance between a pair of monomers $i$ and $j$, $R(|i-j|) = \langle|\vec{x}_i - \vec{x}_j|\rangle$, should increase as their contour distance increases up to its maximum value $|i-j| = N_\mathrm{sub}$. Beyond $|i-j| = N_\mathrm{sub}$, however, the monomers are independent, and their average distance is constant $R_\mathrm{max}(|i-j|) \approx R$ [see Fig.~\ref{fig:monomer_density}(b) inset]. In other words, for any chain length $N$ and a confining sphere radius $R$, the reduced internal distance curves, $R(|i-j|)/R$ vs $|i-j|/N_\mathrm{sub}$, should collapse onto each other. In Fig.~\ref{fig:monomer_density}(b), we verify the independent-chain hypothesis, where we rescale the whole set of internal distance curves shown in the inset with the two different expressions of $N_\mathrm{sub}$: (i) $N_\mathrm{sub} \simeq (D/a)^{1/\nu}$ and (ii) $N_\mathrm{sub}=R^{1 / (3 \nu-1)} N^{(2 \nu-1)/(3 \nu -1)}$. Note that, for weakly confined chains (represented in blue and green), the curves collapse nicely when rescaled by (i) [as well as by (ii) for the the size of chain $N=1000$ we simulated]. As the degree of confinement becomes strong (e.g., red curves), rescaling by (ii) produces a much better result, supporting our argument above. In both cases, we note that the plateau starts at $|i-j|/N_\mathrm{sub} \approx 1$.

Our results have implications for spatial organization of chromosomes inside a eukaryotic cell nucleus. While the eukaryotic chromosomes have several levels of high-order structures, the length scale that characterizes the dsDNA and chromatin fiber is typically $\alt 100$ nm, much smaller than the size of the cell nucleus $\agt 10~\mu$m.  It is thus conceivable that the global organization of these molecules in  confinement will not sensitively reflect molecular details (e.g.,  structure of chromatin fiber), and, importantly, our analysis in Fig.~\ref{fig:stacking_blobs}(b) implies that the free energy cost for global chromosome reorganization is low. 
On the other hand, once territories are formed (by proteins and DNA cross-linkings), the \emph{structured}, compact and segregated chromosomes, are likely to exclude each other, 
since the overlapping of tightly packed blobs (such as star polymers) is highly costly ($\mathcal{F}_\mathrm{n}$) for large $n$, as mentioned earlier.
Further consideration of other factors  including the role of chain stiffness and the geometry of confined space is certainly warranted.    

We thank Daan Frenkel for invaluable discussions, and John Bechhoefer and Rhoda Hawkins for critical reading. This work was in part supported by NSERC (Canada).  SJ acknowledges NSERC post-doctoral fellowship.

\end{document}